\documentclass[aps,prb,twocolumn,amsmath,superscriptaddress]{revtex4-1}

\usepackage{graphicx}

\begin{document}


\title{Thermoelectric effects in superconductor-ferromagnet tunnel junctions on europium sulfide}



\author{S. Kolenda}
\affiliation{Institute of Nanotechnology, Karlsruhe Institute of Technology (KIT), Karlsruhe, Germany}
\author{C. S\"urgers}
\affiliation{Physikalisches Institut, Karlsruhe Institute of Technology (KIT), Karlsruhe, Germany}
\author{G. Fischer}
\affiliation{Physikalisches Institut, Karlsruhe Institute of Technology (KIT), Karlsruhe, Germany}
\author{D. Beckmann}
\email[e-mail address: ]{detlef.beckmann@kit.edu}
\affiliation{Institute of Nanotechnology, Karlsruhe Institute of Technology (KIT), Karlsruhe, Germany}

\date{\today}

\begin{abstract}
We report on large thermoelectric effects in superconductor-ferromagnet tunnel junctions in proximity contact with the ferromagnetic insulator europium sulfide.
The combination of a spin-splitting field and spin-polarized tunnel conductance in these systems breaks the electron-hole symmetry and leads to spin-dependent thermoelectric currents. We show that the exchange splitting induced by the europium sulfide boosts the thermoelectric effect in small applied fields and can therefore eliminate the need to apply large magnetic fields, which might otherwise impede applications in thermometry or cooling.
\end{abstract}

\pacs{74.25.fg,74.40.Gh,74.45.+c,74.78.Na,85.75.-d}

\maketitle


\section{Introduction}

The interplay of the antagonistic quantum mechanical ground states superconductivity and ferromagnetism leads to
new physical effects which form the basis of superconducting spintronics~\cite{linder_superconducting_2015,eschrig_spin-polarized_2015}.
In this context, interesting discoveries in the recent years are the generation of long-range triplet supercurrents~\cite{bergeret_odd_2005,keizer_spin_2006,khaire_observation_2010,robinson_controlled_2010,anwar_long-range_2010} 
in ferromagnetic Josephson junctions as well as the long-range quasiparticle spin-transport~\cite{hubler_long-range_2012,quay_spin_2013} in high-field superconductors. The latter is the result of coupled spin and heat currents~\cite{silaev_long-range_2015,bobkova_long-range_2015,krishtop_nonequilibrium_2015,bobkova_injection_2016}, which give rise to large spin-dependent 
thermoelectric effects in spin-polarized tunnel junctions~\cite{machon_nonlocal_2013,ozaeta_predicted_2014,kalenkov_electron-hole_2014}.
In a previous study~\cite{kolenda_observation_2016,kolenda_nonlinear_2016-1} we demonstrated the existence of thermoelectric currents in a superconductor/ferromagnet tunnel junction experimentally. These thermoelectric effects might be useful for Peltier microrefrigeration \cite{giazotto_opportunities_2006,muhonen_micrometre-scale_2012} as well as high-resolution local thermometry \cite{giazotto_ferromagnetic-insulator-based_2015}. One impediment to these applications is the need to apply large magnetic fields to induce a spin splitting of the density of states. Several proposals have been made to replace the applied field by an intrinsic exchange field provided by the proximity effect with a ferromagnet \cite{giazotto_ferromagnetic-insulator-based_2015,linder2016,bathen2017}.

Here, we expand our studies to superconducting structures which are in proximity contact with the ferromagnetic insulator europium sulfide. 
These systems are well studied~\cite{hao_thin-film_1991,moodera_electron-spin_1988,xiong_spin-resolved_2011,wolf_spin-polarized_2014} and it is known that the quasiparticles in the superconductor are polarized by scattering at the interface to the ferromagnetic insulator, 
so that an additional spin-splitting in the quasiparticle density of states is induced. 
Our goal in this paper is to investigate the influence of this exchange splitting on the generation of the thermoelectric effect and its possible use for eliminating the need of large applied fields.

\section{\label{model} Model}

\begin{figure}
\includegraphics[width=\columnwidth]{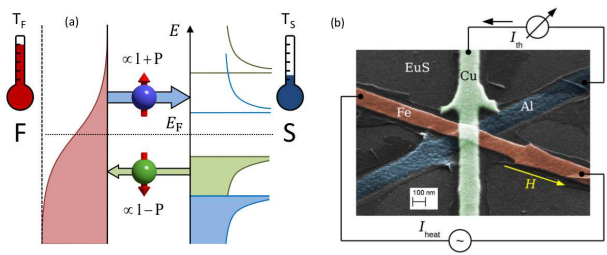}%
 \caption{\label{fig_model_sample}(color online)
(a) Scheme of the generation of thermoelectric current in a superconductor-ferromagnet tunnel junction in the presence of a large spin-splitting field.
(b) False-color scanning electron microscopy image of the central region of sample EUS2 together with the scheme for the thermoelectric measurements.}
 \end{figure}

We start the paper with an introduction to the theory model which we use to describe thermoelectric currents in our systems.
Throughout this work, we use the abbreviations F, I, S, and N to denominate ferromagnetic, insulating, superconducting, and normal metal parts of our structures, e.g., FIS for a 
ferromagnet/insulator/superconductor junction.
Figure~\ref{fig_model_sample}(a) shows schematically how thermoelectric currents are generated across a FIS junction in the presence of a spin-splitting field.
In the superconductor the quasiparticle density of states (DOS) is strongly energy dependent and the 
energies of spin-up and spin-down quasiparticles are shifted by $\pm \mu_\mathrm{B}B_\mathrm{eff}$ with respect to each other.
Here, $\mu_\mathrm{B}$ is the Bohr magneton and
\begin{equation}
 B_\mathrm{eff} = \mu_0H + B^*.
\end{equation}
is the effective spin-splitting field in the superconductor.
It consists of the applied field $\mu_0H$ and the intrinsic exchange field $B^*$ which results from the proximity coupling of S to the ferromagnetic insulator~\cite {hao_thin-film_1991,wolf_spin-polarized_2014}.
The electronic temperature of the ferromagnet is increased by a small thermal excitation to $T_\mathrm{F} = T + \delta T$, while in the superconductor it stays at $T_\mathrm{S} = T$. 
The temperature increase in F leads to an increased population of states with energies above the Fermi level $E_\mathrm{F}$ as well as to an increased number of unoccupied states with energies below $E_\mathrm{F}$.
Once the thermal energy of the electrons in the ferromagnet is large enough, two competing tunnel currents form.
High energy electrons tunnel from the occupied states in F into free states in S, 
while at the same time low energy electrons tunnel from S into the free states of F. 
As indicated in Fig.~\ref{fig_model_sample}(a) these currents are spin-polarized due to the energy dependence and the spin splitting of the quasiparticle DOS by $B_\mathrm{eff}$. 
Both currents would cancel each other out in an unpolarized junction,
but the polarization $P$ of the ferromagnetic junction lifts the symmetry in between them, so that a net tunnel current flows which is driven by the temperature difference $\delta T$.

To model the thermoelectric current across the FIS junction in the presence of a voltage $V$ and a temperature difference $\delta T$ we use equation (2a) of Ref.~\onlinecite{ozaeta_predicted_2014},

\begin{multline}
 I(T,\delta T,V) = \frac{G_\mathrm{T}}{e}\int \left[N_0(E)+{PN_\mathrm{z}(E)}\right]\\ 
\times\left[f_0(E-eV,T+\delta T)-f_0(E,T)\right]dE.
  \label{eqn_Ithermoelectric}
\end{multline}

Here, $G_\mathrm{T}$ is the normal-state tunnel conductance, $e$ is the (negative) charge of the electron, $P = (G_\uparrow-G_\downarrow)/(G_\uparrow+G_\downarrow)$ the polarization of the junction and $f_0$ is the Fermi distribution.
The DOS in the superconductor is split into a symmetric part $N_0 (E) = (N_\uparrow+N_\downarrow)/2$ and an asymmetric part $N_\mathrm{z} (E) = (N_\uparrow-N_\downarrow)/2$
where $N_{\uparrow/\downarrow}$ are the densities of states for spin-up and spin-down quasiparticles respectively.
Both are calculated by the standard model for high-field tunneling~\cite{maki_pauli_1964,meservey_tunneling_1975}.
Equation~(\ref{eqn_Ithermoelectric}) enables us to model the differential conductance $g(V)$ of the tunnel junction in the absence of a thermal excitation $(V\neq 0, \delta T = 0)$ 
as well as the modeling of the thermoelectric current $I_\mathrm{th}$ in the absence of an excitation voltage $(V=0, \delta T\neq 0)$ on an equal footing.
For small voltages and temperature difference it can be linearized to 
\begin{equation}
 I = gV+\eta \frac{\delta T}{\overline{T}},
   \label{eqn_Igeneral}
\end{equation}
where $g$ is the conductance of the junction and $\eta$ is the thermoelectric coefficient while $\overline{T}$ is the average temperature of the junction.
The coefficient $\eta$ is related to the Seebeck coefficient $S = -V/\delta T$, which is widely used to classify thermoelectric effects, by $\eta = Sg\overline{T}$.

\section{\label{samples} Samples and Experiment}


The sample preparation was done in a two stage process. 
First, europium sulfide (EuS) films were deposited on silicon(111) substrates which were heated to temperature $T_\mathrm{sub} \approx 700 - 800 ^\circ \mathrm{C}$
during the evaporation.
X-ray diffractometry showed that under these conditions films which are highly textured in the $\langle 111\rangle$-direction were obtained. 
The magnetization measurements of the film revealed a saturation magnetization $M_\mathrm{S} = 5.3~\mu_\mathrm{B}$ per formula unit and a Curie temperature $T_\mathrm{Curie} = 17.9~\mathrm{K}$. 
A more detailed description of the EuS film deposition and characterization procedure can be found elsewhere~\cite{wolf_fabrication_2014}.
In the second step we fabricated metallic structures on top of the EuS films.
First, PMMA resist was spin coated on the EuS layer and resist masks were structured by means of electron beam lithography.
After a short Ar milling step, a superconducting aluminum wire of thickness $t_\mathrm{Al} \approx 20~\mathrm{nm}$ was evaporated and oxidized in-situ to form a thin insulating tunnel barrier.
Subsequently, ferromagnetic iron ($t_\mathrm{Fe} \approx 20~\mathrm{nm}$) and normal metallic copper ($t_\mathrm{Cu} \approx 50~\mathrm{nm}$) wires were overlaid by shadow evaporation from different directions.
The central part of our sample, the ferromagnetic tunnel junction, is shown in a false-color scanning electron microscopy image in Figure~\ref{fig_model_sample}(b).   
The FIS junction which is formed by the aluminum wire and the iron wire is overlaid by an additional copper wire.
As in our previous experiments~\cite{kolenda_observation_2016}, the samples had an additional normal-metal tunnel junction (not shown here) which was used for control measurements. Here, we show results from three samples labeled EUS1-EUS3, and one sample (FIS1) from our previous study \cite{kolenda_observation_2016} for comparison.

For the measurements, the samples were mounted in a shielded box thermally anchored to the mixing chamber of a dilution refrigerator and cooled down to base temperatures $T_0 = 50 - 500~\mathrm{mK}$.
To avoid confusion, throughout this paper, we use the notation $T_0$ for the bath temperature,
while the electron temperatures of the ferromagnet and of the superconductor are denoted by $T_\mathrm{F}$ and $T_\mathrm{S}$ respectively.
The magnetic field $H$ was applied in the sample plane and parallel to the iron wire as it is indicated by the arrow in Fig.~\ref{fig_model_sample}(b).
The measurements were done in the following order.
First, we measured the local conductance $g = dI/dV$ as a function of the applied bias $V$ with standard lock-in technique to characterize the tunnel junction
(the appropriate measurement scheme is shown schematically in Fig. 1(b) of reference~\onlinecite{kolenda_observation_2016}).
In the next step, we calibrated the temperature difference $\delta T = T_\mathrm{F} - T_\mathrm{S}$ as a function of the applied heater current $I_\mathrm{heat}$ 
by measuring conductance curves $g(V)$ for different values of $I_\mathrm{heat}$. 
Afterwards, we proceeded with the actual thermoelectric measurements. 
A scheme of the measurement configuration is sketched in Figure~\ref{fig_model_sample}(b).
An ac-current $I_\mathrm{heat} = I_0 \sin(\omega t)$ is applied to the iron wire which results in a heating power $P \propto I_\mathrm{heat}^2$. 
This leads to a thermal excitation $\delta T \propto P$ across the junction which is proportional to the second harmonic of the applied frequency $\omega$.
Hence, the thermoelectric current $I_\mathrm{th}$ can be directly monitored in the second harmonic of the resulting current with a lock-in amplifier.

\section{\label{characterization} Characterization and Calibration}

 \begin{figure}
\includegraphics[width=\columnwidth]{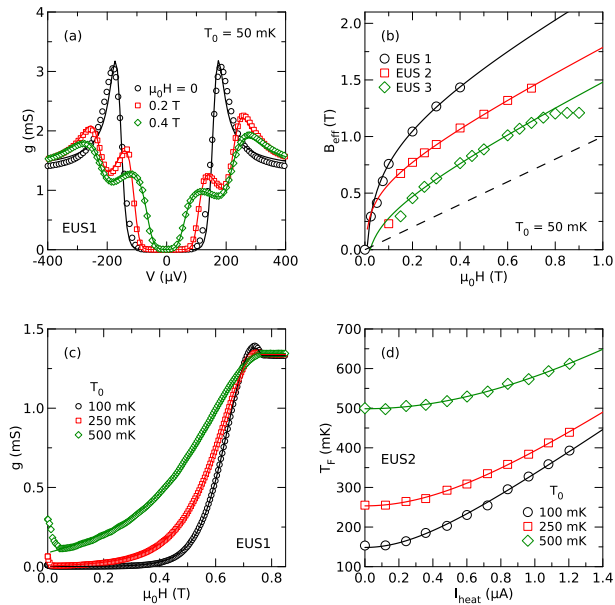}%
 \caption{\label{fig_conductance}
(a) Differential conductance $g$ of the ferromagnetic tunnel junction as a function of the applied bias $V$ for various applied fields $\mu_0H$.
(b) Effective spin-splitting field $B_\mathrm{eff}$ as a function of the applied field $\mu_\mathrm{0} H$ for all samples. The dashed line indicates $\mu_0H$.
(c) Zero-bias conductance $g$ as a function of $\mu_\mathrm{0} H$.
(d) Temperature of the ferromagnet $T_\mathrm{F}$ as a function of the applied heater current $I_\mathrm{heat}$.
}
\end{figure}

We start the discussion of our results with the sample characterization which forms the basis for the analysis of the thermoelectric measurements. 
Figure~\ref{fig_conductance}(a) shows the differential conductance $g$ of the FIS junction of sample EUS1 
as a function of the applied bias $V$ for various applied fields $\mu_0H$ and base temperature $T_0 = 50~\mathrm{mK}$.
At zero field $\mu_0H = 0$, the conductance exhibits the behavior of a high quality tunnel junction with negligible conductance at small voltages and coherence peaks at the gap voltage. Upon increasing the field the coherence peaks split into two peaks for the two spin projections. 
Furthermore, the conductance curves broaden due to the orbital pair breaking effect of the applied field.  
To analyse the effective spin-splitting $B_\mathrm{eff}$, the orbital pair breaking $\Gamma$, the polarization $P$ and the spin orbit scattering strength $b_\mathrm{so}$ we fitted the conductance curves to equation~(\ref{eqn_Ithermoelectric}). 
The fits are plotted in Fig.~\ref{fig_conductance}(a) as solid lines and show good agreement with the data. 
For the spin polarization of the junction we obtained $P \sim 0.15 - 0.2$ which is reasonable for a tunnel junction with a thin insulating layer~\cite{munzenberg_superconductor-ferromagnet_2004}. 
An overview of the sample properties is given in Table~\ref{tab_sampleprop}.

In Figure~\ref{fig_conductance}(b) we show the spin-splitting field $B_\mathrm{eff}$ extracted from the fits of the conductance spectra as a function of the applied magnetic field $\mu_0H$ for all three samples.
The dashed line marks the Zeeman splitting $\mu_0H$ which is expected in the absence of an exchange field. 
$B_\mathrm{eff}$ exceeds $\mu_0H$  value for all three samples and differs from sample to sample.
We attempted to describe the exchange field $B^* = B_\mathrm{eff} - \mu_0H$ with different phenomenological models and obtained the best fits using the logarithmic field dependence $B^* \propto \ln(H)$
which has been reported in the work of Xiong et al.~\cite{xiong_spin-resolved_2011}.
In Fig.~\ref{fig_conductance}(b) we show fits (solid lines) according to 
\begin{equation}
 B_\mathrm{eff} = \mu_0H + a\cdot\ln(H/H_0)
 \label{eqn_Beff}
\end{equation}
with the phenomenological parameters $a$ and $H_0$.
We find sufficient agreement in the field regime $0.1~\mathrm{T} < H < 1~\mathrm{T}$ and use this equation for describing $B_\mathrm{eff}(H)$ for the fits of the field-dependent quantities later on. 
Note, that the field dependence of $B^*$ does not reflect the magnetization of the pure EuS film which had a coercive field $\mu_0H_\mathrm{co} \sim 5~\mathrm{mT}$.
We trace this fact as well as the different strength of the exchange fields for the various samples back to the two step fabrication process.
We assume that the surface of the EuS film is slightly damaged during the argon milling step, leading to variations in the magnetic properties of the Al/EuS interface.

\begin{table}
 \caption{\label{tab_sampleprop}Overview of the sample properties.
 Junction properties: Normal-state tunnel conductance $G_\mathrm{T}$ and spin-polarization $P$.
 Properties of the aluminum wire: critical temperature $T_\mathrm{c}$, critical magnetic field $\mu_0H_\mathrm{c}$, pair potential $\Delta_{0}$ and orbital critical field $\mu_0H_\mathrm{c,orb}$. \hfill}
 \begin{ruledtabular}
 \begin{tabular}{lcccccc}
 Sample    & $G_\mathrm{T}$ & $P$ & $T_\mathrm{c}$ & $\mu_0H_\mathrm{c}$ & $\Delta_{0}$ & $\mu_0H_\mathrm{c,orb}$ \\
   & (mS) &  & (K) & (T) & ($\mu$eV) & (T) \\
 \hline
EUS1 & 1.33 & 0.15 & 1.4 & 0.76 & 1.98 & 1.1 \\
EUS2 & 0.66 & 0.17 & 1.43 & 0.96 & 1.92 & 1.23 \\
EUS3 & 0.78 & 0.185 & 1.47 & 1.01 & 1.90 & 1.16 \\
 \end{tabular}
 \end{ruledtabular}
\end{table}

In Figure~\ref{fig_conductance}(c) we show the zero-bias conductance $g(V=0)$ as a function of $\mu_0H$ for different base temperatures $T_0$.
The conductance exhibits the expected behavior while the critical field $\mu_0H_\mathrm{c}$ where the normal-state conductance $G_\mathrm{T}$ is reached depends on the strength of the exchange field and differs from
sample to sample (see Table~\ref{tab_sampleprop}).
Near $H = 0$, we observed an increased conductance which was not observed for the samples without EuS substrates~\cite{kolenda_observation_2016}. We attribute this to the inhomogeneous magnetization of the EuS film in small fields which may induce either an out-of-plane stray field or an inhomogeneous exchange splitting and therefore weaken superconductivity. 

For the fits of $g(H)$ we assume that the orbital pair-breaking strength $\Gamma$ follows the dependence~\cite{maki_pauli_1964}
\begin{equation}
 \frac{\Gamma}{\Delta_0} = \frac{1}{2} \left(\frac{H}{H_\mathrm{c,orb}}\right)^2
\end{equation}
for a thin film with in-plane magnetic field, where $\Delta_0 = \Delta(T=0,H=0)$ is the pair potential at zero temperature and zero field and $H_\mathrm{c,orb}$  is the orbital critical field in the absence of Zeeman splitting. $\Delta(T,H)$ and $B_\mathrm{eff}(T,H)$ were calculated self-consistently using the model of Alexander et al.~\cite{alexander_theory_1985}, which includes Fermi-liquid renormalization of the spin splitting with the Fermi-liquid renormalization parameter $G_0$. We follow Ref.~\onlinecite{xiong_spin-resolved_2011} and apply the renormalization to the effective field $B_\mathrm{eff}$ modeled by eq.~(\ref{eqn_Beff}).

By fitting the zero-bias conductance according to Eq.~(\ref{eqn_Ithermoelectric}) we extracted the remaining junction parameters ($G_\mathrm{T}$, $\Delta_{0}$ and $H_\mathrm{c,orb}$).
Fit parameters extracted from the fits at $T_0 = 100~\mathrm{mK}$ are given in Table~\ref{tab_sampleprop} for all samples. 
For $G_0$ we obtained $G_0 = 0.21$ (EUS1) and $G_0 = 0.26$ (EUS2 and EUS3) which are in reasonable agreement with literature values for thin aluminum films~\cite{tedrow_spin-polarized_1984,catelani_fermi-liquid_2008}.

For the calibration of the thermal excitation $\delta T$ across the junction, we applied a dc heater current $I_\mathrm{heat}$ to the iron wire and measured the differential conductance $g(V)$ to obtain the electron temperature $T_\mathrm{F}$ as a function of $I_\mathrm{heat}$.
Details of the procedure can be found in Ref.~\onlinecite{kolenda_observation_2016}. 
The results of these measurements are shown in Figure~\ref{fig_conductance}(d) for different base temperatures $T_0$ using the example of sample EUS2. $T_\mathrm{F}$ increases monotonically with increasing  $I_\mathrm{heat}$. At low temperatures, $T_\mathrm{F}$ is slighly larger than $T_0$ even for $I_\mathrm{heat}=0$, which we attribute to incomplete filtering of the measurement lines.
Solid lines show fits to the model of a mesoscopic wire in quasi-equilibrium with negligible electron-phonon scattering~\cite{giazotto_opportunities_2006}
\begin{equation}
 T_\mathrm{F}=\sqrt{T^2+\frac{I_\mathrm{heat}^2R_\mathrm{heat}^2}{4L_0}},
   \label{eqn_heatmodel}
\end{equation}
where $R_\mathrm{heat}$ is the resistance of the heater wire, $L_0=\pi^2k_\mathrm{B}^2/3e^2$ is the Lorenz number and $T$ is the electron temperature in the absence of heating.
$R_\mathrm{heat}$ was left as a free fit parameter and was usually found to be of the same order, but smaller than the two-probe resistance of the iron wire. We attribute this to the thick copper wire on top of the junction which acts as a cooling fin. The fits are in good agreement with the data, and we use them to estimate the temperature difference $\delta T_\mathrm{cal}(I_\mathrm{heat})$ which is generated across the junction for a certain heater current. 

\section{\label{thermoelectric} Thermoelectric current}

\begin{figure}
\includegraphics[width=\columnwidth]{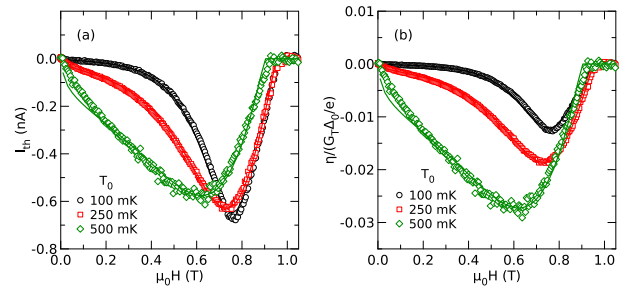}%
 \caption{\label{fig_TEcurrent}
(a) Thermoelectric current $I_\mathrm{th}$ measured for sample EUS2 as a function of the applied magnetic field $\mu_0H$ for different base temperatures $T_0$.
(b) Normalized thermoelectric coefficient $\eta$, inferred from the data of panel (a) for different base temperatures $T_0$. 
In both panels solid lines show fits of the thermoelectric current according to Eq.~(\ref{eqn_Ithermoelectric}).
}
\end{figure}

We now turn to the main results of this paper, the thermoelectric measurements with $V=0$ and $\delta T\neq 0$.
Figure~\ref{fig_TEcurrent}(a) shows the thermoelectric current $I_\mathrm{th}$ as a function of the applied magnetic field $\mu_0H$ for different base temperatures $T_0$. 
For the measurement of each curve, we adjusted the applied thermal excitation to $\delta T_\mathrm{cal} \approx 100~\mathrm{mK}$. 
The thermoelectric current is always negative (corresponding to electrons tunneling into the superconductor), and in the following discussion we refer to its magnitude.
For the lowest temperature, $T_0 = 100 ~\mathrm{mK}$, there is no signal at zero field, and $|I_\mathrm{th}|$ starts to rise for $\mu_0H \approx 0.3~\mathrm{T}$.
Then, the current grows with increasing spin-splitting of the quasiparticle DOS until the spectral gap closes.
At this point, $|I_\mathrm{th}|$ exhibits its maximum. 
Above this field, the current decreases again and vanishes finally as $H$ approaches the critical field $H_\mathrm{c}$. 
For higher base temperatures the current grows faster in small fields, and the maximum is broadened.

The solid lines in Figure~\ref{fig_TEcurrent}(a) indicate the fits of the thermoelectric current to the theory model.
We used equation~(\ref{eqn_Ithermoelectric}) together with the fit parameters extracted from the fits of $g(H)$.
Only the thermal excitation $\delta T_\mathrm{fit}$ was left as free fit parameter. 
We find $\delta T_\mathrm{fit} \sim 0.8 - 0.9 \delta T_\mathrm{cal}$, with good agreement between the data and the fits for all base temperatures. 
The reduction of the fitted thermal excitation compared to the calibration value can be attributed to indirect heating of the superconductor via the thermal conductance of the junction.
Control experiments which were done in analogy to our previous work (see Fig. 3 in Ref.~\onlinecite{kolenda_observation_2016})
revealed that the electronic temperature of the superconductor $T_\mathrm{S}$ increases by about $10-20\%$ of the thermal excitation applied to the iron wire. 

Figure~\ref{fig_TEcurrent}(b) shows the thermoelectric coefficient $\eta = I_\mathrm{th}\cdot (\overline{T}/\delta T)$ as a function of the applied field $\mu_0H$.
It is inferred from the raw data in panel (a) and normalized to the normal state conductance $G_\mathrm{T}$ and the pair potential $\Delta_0/e$ to make it dimensionless and comparable.
For the thermal excitation we use $\delta T = \delta T_\mathrm{fit}$.
The behavior of $\eta$ is found in agreement with our previous results~\cite{kolenda_observation_2016} as well as it follows the theoretical prediction~\cite{ozaeta_predicted_2014}.

\begin{figure}
\includegraphics[width=\columnwidth]{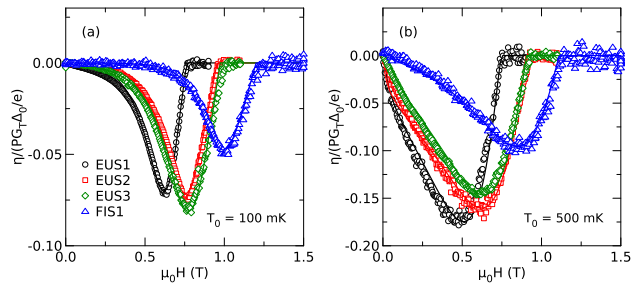}%
 \caption{\label{fig_eta_compare}
Thermoelectric coefficient $\eta$ normalized to $P$, $G_\mathrm{T}$ and $\Delta_0/e$ as a function of the applied magnetic field $\mu_0H$ for different samples.
The data are for base temperature $T_0 = 100~\mathrm{mK}$ (a) and $T_0 = 500~\mathrm{mK}$ (b), respectively.}
\end{figure}

Finally, we compare the thermoelectric signal for all three samples with each other and with the sample FIS1 from our previous work~\cite{kolenda_observation_2016}.
Sample FIS1 has the same sample layout, but was structured on top of SiO${}_2$ without an EuS film below the superconductor.
Figure~\ref{fig_eta_compare} shows the normalized thermoelectric coefficient $\eta$ as a function of $\mu_0H$ for the base temperatures $T_0 = 100~\mathrm{mK}$ and $T_0 = 500~\mathrm{mK}$.
To obtain comparable signals, $\eta$ was additionally normalized to the junction polarization $P$ here. 
We observe that for both base temperatures and all samples the qualitative behavior of the thermoelectric signal is similar. 
However, for the samples EUS1, EUS2, and EUS3 with exchange field, the overall signal amplitude is much larger, and the onset of the thermoelectric signal is shifted to lower fields compared to the sample FIS1 without an EuS film. In particular at $T_0 = 500~\mathrm{mK}$, the thermoelectric effect is increased considerably in small fields. 

\section{Conclusion}

In conclusion, we have shown the influence of the intrinsic exchange field on thermoelectric currents in superconductor-ferromagnet tunnel junctions on top of europium sulfide films. 
The overall magnitude and field dependence of the thermoelectric current is similar to our previous study on structures without exchange splitting. Due to the increase of the effective spin-splitting by the exchange field, the thermoelectric currents are larger and appear at smaller magnetic fields for the structures on top of europium sulfide films.
Hence, the use of proximity coupling of the superconductor with a ferromagnetic insulator can eliminate the need to apply large magnetic fields for the generation of thermoelectric currents. These structures are a further step towards improved thermoelectric low-temperature devices and might enable high-resolution thermometry and efficient microrefrigeration.

\begin{acknowledgments}
This work was partially supported by the DFG under grant number BE-4422/2-1.
\end{acknowledgments}

\bibliography{TE_on_EuS.bib}

\end{document}